\documentclass[12pt]{article}
\usepackage{amsfonts}
\usepackage{amsmath}
\usepackage{amssymb}
\usepackage{authblk}
\usepackage[portrait, margin=1in]{geometry}
\usepackage{graphicx}
\usepackage{hyperref}
\usepackage{setspace}
\usepackage[normalem]{ulem}
\title{\footnote{Essay written for the Gravity Research Foundation 2026 Awards for Essays on Gravitation.}~Black hole mergers as probes of spacetime's\\condensed degrees of freedom?}

\author{Arno~Keppens$^1$~(\href{mailto:arno.keppens@spacepole.be}{arno.keppens@spacepole.be}, corresponding author) Lester~Kurvers$^2$~(\href{mailto:lester.kurvers@ugent.be}{lester.kurvers@ugent.be})\\$^1$Space Pole, 1180 Brussels, Belgium\\$^2$Dept.~of Physics and Astronomy, Ghent University, 9000 Ghent, Belgium}

\date{March 30, 2026}

\begin{document}
\maketitle
\begin{abstract} 
Black hole physics currently lacks a fully coherent understanding of the black hole mass (density), entropy, and interior (non-)singularity. These concepts are related to the black hole radius, area (of the horizon), and volume (within the horizon), respectively, in the Schwarzschild solution to Einstein's field equations. In this work, we argue that these concepts can be given reasonable interpretations in terms of spacetime's thermodynamic degrees of freedom, which constitute the metric, when the black hole is considered as a condensate thereof. Recent observations of black hole merger events support our proposal.
\end{abstract}

\newpage
\section*{Theory}

Spherically symmetric solutions to Einstein's field equations have a metric line element of the form:
\begin{equation}\label{sss}
    ds^2 = -f(r) (c dt)^2 + f(r)^{-1} dr^2 + r^2 d\Omega^2
\end{equation}
with $f(r)$ a function of the radius $r$ with respect to the center of symmetry, $c$ the invariant speed of light making $c dt$ a spatial differential, and $\Omega$ capturing the angular dimensions.
A first analytical expression for $f(r)$ was provided by Schwarzschild for the gravitational field surrounding a spherical mass $M$~\cite{Schwarzschild-1916}:
\begin{equation}\label{S}
    f_S(r) = \left( 1 - \frac{r_S}{r} \right)
\end{equation}
with $r_S = 2 G M / c^2$ called the Schwarzschild radius, including the gravitational constant $G$. 
This Schwarzschild solution however features an intrinsic metric curvature singularity for $r = 0$, which is problematic for the physical interpretation of the Schwarzschild black hole. The Schwarzschild--de Sitter black hole was therefore proposed as an alternative by Dymnikova~\cite{Dymnikova-1996}, where the central Schwarzschild singularity is overwritten by an overall flat de Sitter metric:
\begin{equation}\label{D}
    f_D(r) = \left( 1 - \frac{r_S}{r} - \frac{r^2 \Lambda}{3}\right)
\end{equation}
given a (cosmological) constant $\Lambda$. The major drawback of this solution is that it introduces a curvature discontinuity at the Schwarzschild radius. In order to arrive at a black hole metric that is both continuous and complete, a gravastar-like metric is required, which comprises an interior de Sitter geometry and an exterior Schwarzschild geometry, separated by a transition shell \cite{Mazur-2004}. The corresponding metric function $f_G(r)$ thus consists of (at least) three partial expressions. Here, however, inspired mainly by \cite{Chapline-2001}, \cite{Visser-2004}, and \cite{Cadoni-2022}, we propose to explicitly extrapolate the Schwarzschild solution at $r = r_S$ to the entire black hole interior, which can hence be considered as a spacetime condensate with constant critical curvature $\Lambda_C$:
\begin{align}
    f_C(r \geq r_S) &= \left( 1 - \frac{2 G M}{c^2 r} \right) \\
    f_C(r \leq r_S) &= \left( 1 - \frac{r^2 \Lambda_C}{3}\right)
\end{align}
Matching these two expressions at $r = r_S$ confirms that the uniform yet critical energy density within the black hole, $c^4 \Lambda_C / 8 \pi G$, is then given by $M c^2 / (4/3) \pi r_S^3$.

We have thus evolved from a pure Schwarzschild solution ($f_S$), to the addition of a flat background ($f_D$), to confining the background within the Schwarzschild horizon ($f_G$), and to saturating the confined background to the Schwarzschild horizon's energy density ($f_C$) -- see \cite{Carballo-2025} and references therein for a broader overview. The last solution is largely motivated by the thermodynamics and fluid dynamics of spacetime \cite{Padmanabhan-2015}: We consider the black hole as a condensate of the constituting atoms of spacetime, while at the same time making our analysis independent of the physical nature of these thermodynamic degrees of freedom. This differs from the approach by Dvali and Gomez that explicitly regards the black hole as a graviton condensate \cite{Dvali-2014}.

The constituent condensate picture of a black hole is demonstrated in Figure 1, which shows the radial spacetime constituent density distributions $\rho$ for three bodies $M_1$ to $M_3$. As the gravitational fields in terms of quadratically damping degrees of freedom densities outside of these bodies (black curved lines) perfectly overlap, the three bodies are experienced to have exactly the same Newtonian mass $M$ by a distant observer (wherefrom the Newtonian point mass concept). Correspondingly, upon condensation of the low-density body $M_1$ to the body $M_3$ at constant critical density $\rho_C$, as a thermodynamic phase transition, the number of spacetime degrees of freedom is conserved, while the entropy must (locally) drop significantly. Indeed, according to classical thermodynamics, the condensed degrees of freedom do no longer contribute to the system's entropy, except for those on the (Schwarzschild) surface as the physical boundary of the condensate, hence the area law \cite{Hawking-1974}.

\begin{figure}[ht]
\centering
\includegraphics[width=0.8\textwidth,trim={0 4cm 0 3cm},clip]{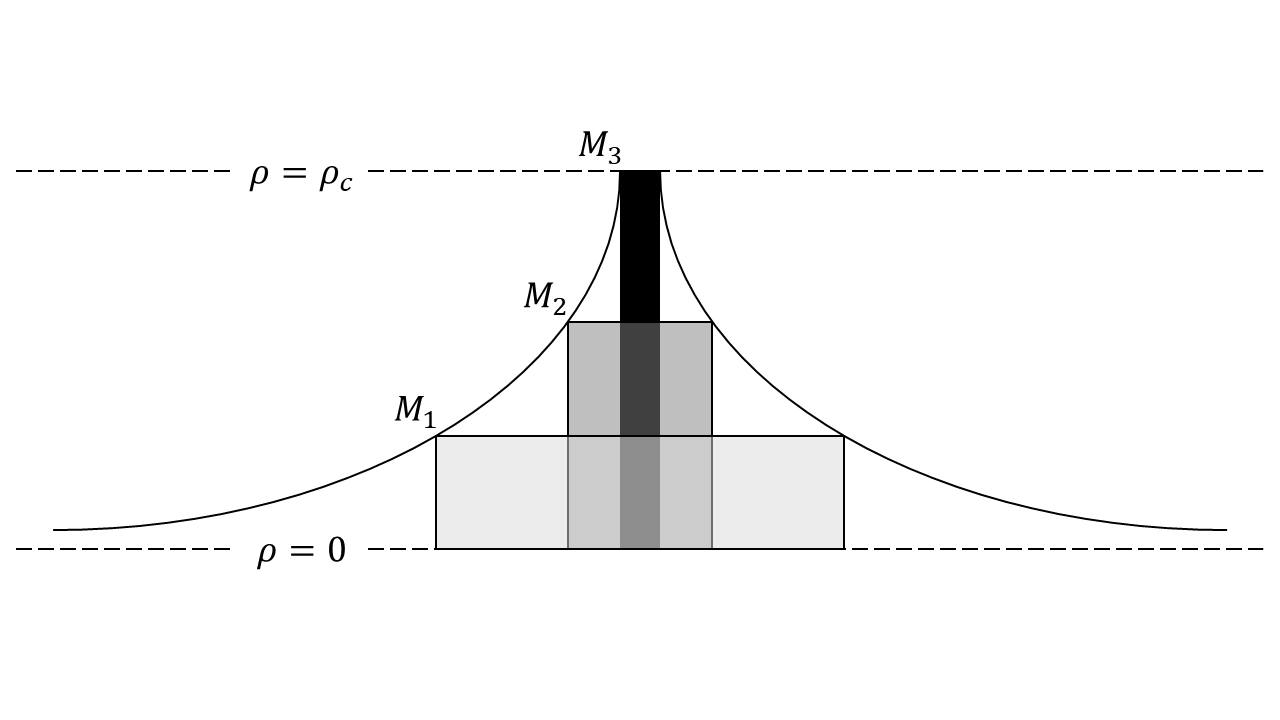}
\caption{Three massive bodies with different degrees of freedom densities $\rho$ are nevertheless experienced to have the same Newtonian mass $M = M_1 = M_2 = M_3$ to a distant observer, because of the masses' overlapping far-field behavior (curved lines). The degrees of freedom are saturated to the critical density $\rho_C$ for the condensate $M_3$.}
\label{fig_masses}
\end{figure}
 
According to Schwarzschild spacetime thermodynamics, equilibrium states, and hence the laws of thermodynamics, exclusively hold for surfaces that are concentric with the Schwarzschild horizon ($r \geq r_S$) -- see \cite{Keppens-2025} and references therein. The number of spacetime degrees of freedom therefore only equals the entropy at these surfaces, up to a constant. For the number of degrees of freedom on the Schwarzschild horizon, we have from the area law: $N_A = S_{BH} / k_B = A_S / (2 l_P)^2$, with $A_S$ the horizon area and $l_P$ the Planck length. Therefrom, given the constant degrees of freedom density within the black hole condensate, $N_R = R_S / (2 l_P)$ and $N_V = V_S / (2 l_P)^3$ for its radial and volumetric constituent numbers, respectively. This entails that the critical density equals one spacetime degree of freedom per cubed double Planck length for all black holes, which contrasts the variable and hence rather misleading black hole mass density concept: Only a Planckion reaches the (modified) Planck mass density of $m_P/(2 l_P)^3$. However, from the Planck mass ($m_P = c^2 l_P / G$) and Schwarzschild radius definitions, one also immediately has $N_R = M / m_P$. This means that Newtonian mass is merely a measure of the radial number of degrees of freedom that uniquely defines a (non-rotating) black hole condensate (or a lower-density body with the same far-field metric, again see Figure 1). The existence of charged black holes is impossible within this picture, because of the already saturated number of degrees of freedom -- see \cite{Pavon-2020} for additional motivation.

Viewing black holes as spacetime constituent condensates also alters our understanding of black hole merging events. It's immediately clear that purely Newtonian mass conservation cannot hold: If two irreducible black holes with equal masses $M \propto r_S$ coalesce, the resulting entropy would be double the sum of the initial entropies, see Table 1. And applying this non-relativistic mass conservation to the merging of condensed black holes would result in a fourfold increase of the total number of condensed spacetime degrees of freedom. This does not make sense (cf.~strike-through in Table 1). What does make sense here is to replace the misleading -- because radial -- Newtonian mass conservation concept for black hole mergers by volumetric degrees of freedom conservation, again see Table 1. In that case, if the end product of the merging were again a static condensed Schwarzschild black hole, the resulting black hole's Newtonian mass $M'$ and entropy $S'$ would be reduced by about 40 and 20 \%, respectively, with respect to the initial combined masses and entropies. For condensates, one indeed expects a stronger (local) entropy reduction for larger bodies, while obviously the total entropy must increase in agreement with the Generalized Second Law of black hole thermodynamics (analogous to its application to that well-known other surface-reducing effect called Hawking radiation) \cite{Bekenstein-1974}. Similarly, total relativistic mass conservation must hold for the merger as a whole and, according to the four laws of black hole mechanics, this implies an area increase \cite{Bardeen-1973}. Assuming the merging results in an effective Kerr black hole of mass $M^* = 2M$, its outer surface area is given by:
\begin{equation}\label{A}
    A^* = 8 \pi \left( \frac{2 G M}{c^2} \right)^2 \left( 1 + \sqrt{1 - a^2} \right)
\end{equation}
for a dimensionless spin $a = cJ/GM^2$. With values of $a$ crowding around 0.7 \cite{Ruppeiner-2026}, one obtains $A^* \approx 1.7 (2A)$. This outer surface expansion result, due to spin from energy conservation, is expected to coexist with a condensed inner core of Newtonian mass $M'$ and entropy $S'$ in agreement with degrees of freedom conservation.

\begin{table}[ht]
\centering
\caption{Expressions for the mass, entropy, and degrees of freedom (d.o.f.)~for the end product of the merging of two equal black hole condensates, in terms of their initial values, differentiated between mass conservation and degrees of freedom conservation.}
\label{table_ensembles}
\begin{tabular}{ccc}
\hline
 & \textbf{$M$ conserved} & \textbf{$N$ conserved}\\
\hline
Mass $M \propto r_S$ & $M^* = 2M$ & $M' = 2^{-2/3} (2M)$\\
Entropy $S \propto A_S$ & \sout{$S^* = 2(2S)$} & $S' = 2^{-1/3} (2S)$\\
D.o.f.~$N \propto V_S$ & \sout{$N^* = 4(2N)$} & $N' = 2N$\\
\hline
\end{tabular}
\end{table}

\newpage
\section*{Observations}

Several observations support our view on black holes as condensates of spacetime's degrees of freedom. First, charged black holes (or their mergers) have not been observed. Current astrophysical evidence thus suggests that real-world black holes are essentially electrically neutral, in agreement with the conclusion above that charged black hole condensates do not exist due to degrees of freedom saturation. On the other hand, the observation of a charged black hole (merger) would immediately falsify the non-singular black hole structure presented here.

Second, within the non-singular black hole paradigm (which sensibly avoids mathematical artifacts being part of physical reality), recent observational merger data clearly support the black hole condensate model $f_C$ in disfavor of the gravastar model $f_G$, both in the ringdown and echoes of black hole merging events: LVK (LIGO Scientific, Virgo, and KAGRA Collaborations) has shown no statistically significant evidence for so-called merger echoes \cite{GWTC4-2025}, which are expected to be produced by gravitational waves that are bouncing between a merged gravastar surface and its surrounding photon sphere. Only if the gravastar surface coincides (within $l_P$) with a Schwarzschild horizon, as in the condensate model proposed here, then gravitational waves are fully absorbed by that horizon and wave pulses following a main merger signal do not occur. LVK moreover observed multiple overtones in the loudest gravitational wave signal from a merger and ringdown event to date \cite{GW250114-2025}. These tones leave no room for the regular gravastar model that should show distinct harmonics as determined by the nature of its exotic transition shell.

Third, following up on others, the latter experiment also confirmed Hawking’s area law. As the observed merger event involved two black holes of near-equal masses and small spins, its initial conditions are close to what has been discussed in the previous section (a merging of equal Schwarzschild black hole condensates). The ratio of the final and initial total areas that was obtained from the measurement and corresponding simulations amounts to 1.6--1.7, in remarkable agreement with our theoretical result. The (order of percentages) fraction of the energy that is dissipated as gravitational waves is not taken into account here.

\section*{Discussion}

This work is situated within the emergent quantum gravity framework, which poses that spacetime and gravity are not fundamental, but are collective, thermodynamic phenomena arising from pre-geometric degrees of freedom that constitute an effective metric. The finite size of these ``atoms'' of spacetime immediately imposes a maximum degrees of freedom density. We therefore conclude that black holes as understood in terms of spacetime singularities are mathematical artifacts that cannot physically exist, and this without the need for modifications of general relativity theory or black hole thermodynamics, as also motivated in \cite{Mathur-2023}. Instead, a black hole forms when the spacetime fluid, due to self-interaction of its constituents, condenses into a spherical solid with maximal degrees of freedom density. As a result, the condensed degrees of freedom do no longer contribute to its entropy, except for those on the (Schwarzschild) surface as the physical boundary of the condensate, yielding the area law. The maximum degrees of freedom density $(2 l_P)^{-3}$ of the black hole condensate also explains why the Unruh temperature effect only correctly follows from the gravitational acceleration on the Schwarzschild surface \cite{Keppens-2025}: Quantum field theories typically (implicitly) assign one degree of freedom to each Planck volume, while within the presented picture this only holds for spacetime condensates.

\newpage
In this work, we argue that irregularities in the thermodynamic quantities upon singular black hole formation and merging point at the possibility of a different picture. Within the non-singular black hole paradigm, which is considered to be more realistic, black hole condensates form and merge in agreement with both degrees of freedom conservation and total mass-energy conservation. Black hole condensates are hence found to be our best non-singular black hole candidates in agreement with recent gravitational wave observations, in contrast with the initial gravastar picture that by now seems to be falsified. The proposed framework in terms of finite spacetime constituents additionally provides the theoretical foundations for unified simulations of entire gravitational systems, which explicitly rely on degrees of freedom conservation \cite{Keppens-2023}. This thermodynamic completeness backs our framework's closeness to reality.


\bibliographystyle{unsrt}
\bibliography{bibtexfile}

\end{document}